\documentclass{iau}

\usepackage{booktabs}
\usepackage{graphicx}
\usepackage{microtype}
\usepackage{hyperref}

\title[Black Holes in Globular Clusters] 
{Stellar-Mass Black Holes in Globular Clusters:\\ Dynamical Consequences and Observational Signatures}

\author[A. Askar, M. Giersz, M. Arca Sedda, A. Askar, M. Pasquato \& A. Leveque ]   
{Abbas Askar$^1$, 
Mirek Giersz$^2$,
Manuel Arca Sedda$^3$,
Ammar Askar$^4$,
Mario Pasquato$^5$
 \and Agostino Leveque$^2$}

\affiliation{$^1$Lund Observatory, Department of Astronomy,\\ and Theoretical Physics, Lund University, Box 43, SE-221 00 Lund, Sweden \\ email: {\tt askar@astro.lu.se} \\[\affilskip]
$^2$Nicolaus Copernicus Astronomical Center, Polish Academy of Sciences,\\
ul. Bartycka 18, 00-716 Warsaw, Poland \\email (MG): {\tt mig@camk.edu.pl}, email (AL): {\tt agostino@camk.edu.pl}\\[\affilskip]
$^3$Astronomisches Rechen-Institut, Zentrum f{\"u}r Astronomie, University of Heidelberg,\\M{\"o}nchhofstrasse 12-14, 69120, Heidelberg, Germany \\ email: {\tt m.arcasedda@gmail.com} \\[\affilskip]
$^4$School of Computer Science, College of Computing, Georgia Institute of Technology,\\ 801 Atlantic Dr, Atlanta, GA 30332, USA\\ email: {\tt aaskar3@gatech.edu} \\[\affilskip]
$^5$INAF, Osservatorio Astronomico di Padova,\\ vicolo dell'Osservatorio 5, 35122 Padova, Italy \\ email: {\tt mario.pasquato@inaf.it} \\[\affilskip] }

\pubyear{2019}
\volume{351}  
\jname{Star Clusters: From the Milky Way to the Early Universe}
\editors{A. Bragaglia, M.B. Davies, A. Sills \& E. Vesperini, eds.}
\begin{document}

\maketitle

\begin{abstract}
Sizeable number of stellar-mass black holes (BHs) in globular clusters (GCs) can strongly influence the dynamical evolution and observational properties of their host cluster. Using results from a large set of numerical simulations, we identify the key ingredients needed to sustain a sizeable population of BHs in GCs up to a Hubble time. We find that while BH natal kick prescriptions are essential in determining the initial retention fraction of BHs in GCs, the long-term survival of BHs is determined by the size, initial central density and half-mass relaxation time of the GC. Simulated GC models that contain many BHs are characterized by relatively low central surface brightness, large half-light and core radii values. We also discuss novel ways to compare simulated results with available observational data to identify GCs that are most likely to contain many BHs.
\keywords{stellar dynamics, globular clusters: general, stars: black holes}
\end{abstract}

\firstsection 
\section{Introduction}

Over the past decade, several stellar-mass black hole (BH) candidates in binary systems have been detected in Galactic globular clusters (GCs) through electromagnetic radiation \cite[(Strader et al. 2012, Bahramian et al. 2017)]{2012Natur.490...71S,2017MNRAS.467.2199B} and radial velocity variations \cite[(Giesers et al. 2018)]{2018MNRAS.475L..15G}. BHs should form in GCs within the first few tens of millions of years from the evolution of massive stars. The number of BH progenitors in a GC depend on its size, initial mass function of stars and metallicity. With a typical IMF, a sample of 1000 stars would contain about 2 - 3 BH progenitors; hence a GC with a million initial stars could produce a few 1000 BHs. If these BHs receive a natal kick with a velocity larger than the escape velocity of the GC, they will not be retained. Conversely, if they receive a natal kick smaller than the GC's escape velocity, they are retained. The escape velocity of a GC at the time when BHs form depends on the GC's initial central concentration and size. The initial parameters of GCs are uncertain and initial escape velocities could vary from few tens to up to a few hundred km/s. Additionally, the exact mechanism responsible for BH natal kicks is unknown and the magnitude of this kick is only weakly constrained. Given these uncertainties, it is difficult to assess the exact retention fraction of BHs in GCs following their formation.

\begin{figure}[t]
\begin{center}
\includegraphics[width=2.1in]{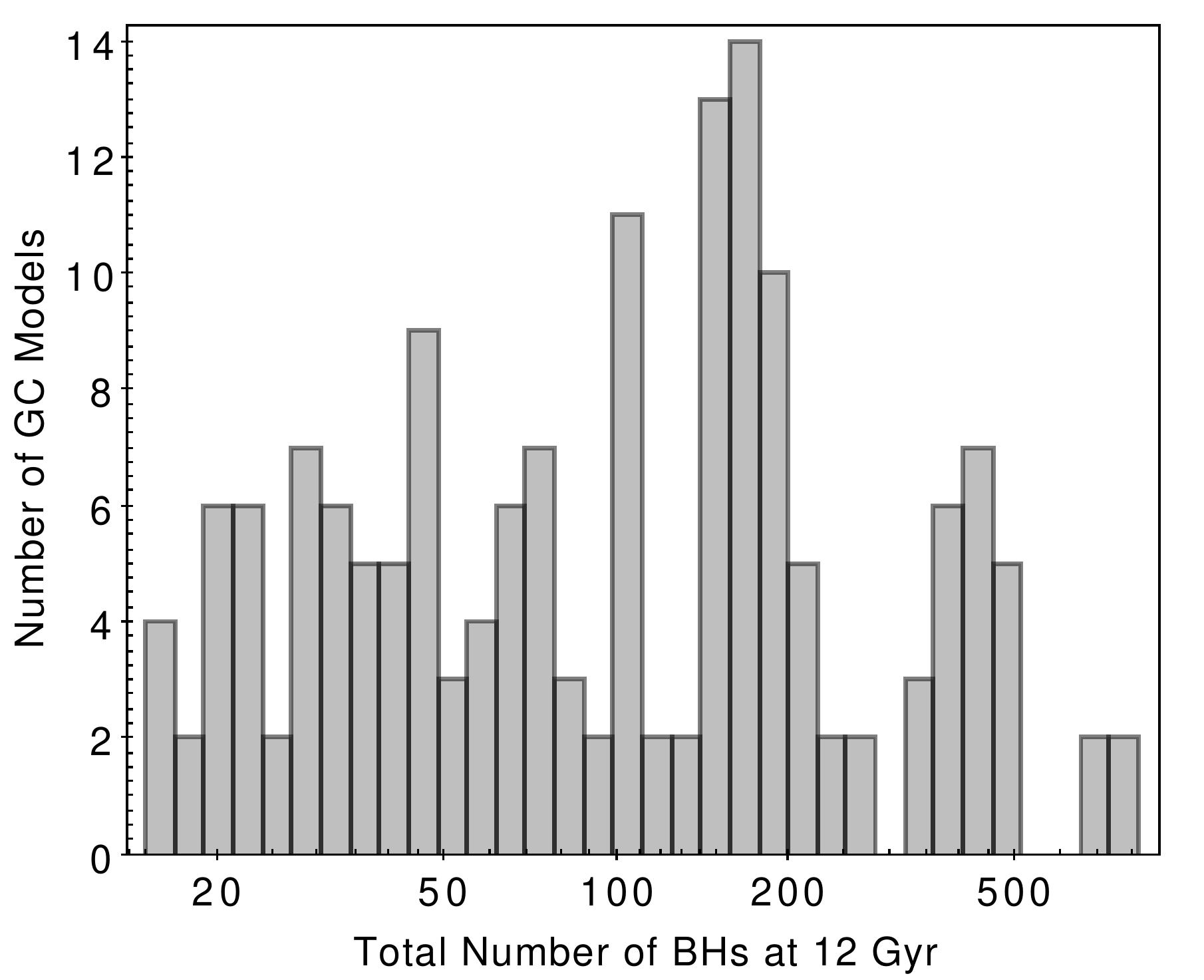} 
 \caption{Distribution of the total number of BHs at 12 Gyr for MOCCA-Survey Database I models with more than 15 BHs at 12 Gyr.}
   \label{fig1}
\end{center}
\end{figure}

Even with a high BH retention fraction during GC formation, the dynamics and long-term survival (up to a Hubble time) of the BH population are not guaranteed. Analytic studies (\cite[Kularni et al. 1993; Sigurdsson \& Hernquist 1993]{1993Natur.364..421K,1993Natur.364..423S}) had suggested that BHs in a GC would quickly segregate and form a subcluster that has a very short relaxation time. Strong dynamical interactions of BHs in this subcluster would lead to their prompt ($\sim 100$ Myr) ejection and only 1 or 2 BHs could survive in a GC up to a Hubble time. However, more recent numerical simulations \cite[(see Morscher et al. (2015) and Giersz et al. (2019),  and references therein)]{2015ApJ...800....9M} and theoretical arguments \cite[(Breen \& Heggie 2013)]{2013MNRAS.436..584B} have shown that BHs can survive in a GC for much longer than previously thought. The evolution of a BH subcluster is governed by the longer two-body relaxation time-scale in the bulk of the GC. Therefore, initially large GCs with long relaxation times may still contain a large number of BHs up to a Hubble time. In this proceeding, we discuss which initial properties of GCs affect how many BHs they might contain after 10-13 Gyr of dynamical evolution. We also highlight recent results that compare simulated GC models with Galactic GCs to identify which Galactic GCs are most likely to contain a substantial number of BHs. 

\section{Simulated Globular Cluster Models}

We closely examined results from nearly 2000 GC simulations simulated using the MOCCA Monte Carlo \cite[(Hypki \& Giersz 2013)]{2013MNRAS.429.1221H} code for star cluster simulation as part of the MOCCA-Survey Database I \cite[(see Askar et al. (2017) and refernces therein for details)]{2017MNRAS.464L..36A} project. About 160 models had more than 15 BHs at Hubble time. The distribution of the number of BHs in these 160 GC models at 12 Gyr is shown in \autoref{fig1}. For these models, the BH natal kicks were modified according to the fallback prescription provided by \cite[Belczynski et al.  (2002)]{2002ApJ...572..407B}. These prescriptions reduce the natal kick received by the BHs, increasing the BH retention fraction in the GCs after their formation. The post formation BH retention fraction in these models was between 15 to 50 per cent depending on other GC initial properties \cite[(see Askar et al. (2018) for details)]{2018MNRAS.478.1844A}. About 85 per cent of the 160 models had initial number of stars larger than $7\times10^{5}$.  

\section{Black Hole Subsystem Size and Globular Cluster Properties}

\cite[Arca Sedda et al. (2018); Kremer et al. (2018); Askar et al. (2018)]{2018MNRAS.479.4652A,2018ApJ...855L..15K,2018MNRAS.478.1844A} found that GC models that keep the most number of BHs at 12 Gyr are dynamically younger with initial half-mass relaxation times longer than $\rm \sim 1 Gyr$. In such GCs, the BH subsystem (BHS) can remain in balanced evolution with the bulk cluster for a longer duration. In the left panel in Fig. 2, we show the total number of BHs over time inside the GC for models with initially $7\times10^{5}$ objects, 10 percent initial binary fraction, metallicty (Z) of $\rm 0.05 \ Z_{\odot}$ and BH natal kicks computed using \cite[Belczynski et al. (2002)]{2002ApJ...572..407B}. The models shown in Fig. 2 have different half-mass radius, tidal radius and central concentration. The right panel of the Fig. 2 shows the evolution of the average BH mass in the GC.

We find that in order to sustain a sizeable BHS up to a Hubble time ($\rm N_{BH} \gtrsim 25$), the GC needs to have a long half-mass relaxation time ($\gtrsim 500$ Myr) and low initial central density ($\rm < 10^{5} \ M_{\odot} \ pc^{-3}$) initially. In all models, the average BH mass always decreases as the most massive BHs in the GC are preferentially ejected via strong interactions at the center of the BHS. \cite[Arca Sedda et al. (2018)]{2018MNRAS.479.4652A} defined the size of the BHS as the radius within which half of the mass is in BHs and the remaining half is in other stars. By this definition, \cite[Arca Sedda et al. (2018)]{2018MNRAS.479.4652A} were able to find correlations between properties of the BHS and the host GC in 160 analysed models. Dynamically younger clusters have more extended and massive BHS with larger average BH Mass.

\begin{figure}[t]
\begin{center}
\includegraphics[width=5.2in]{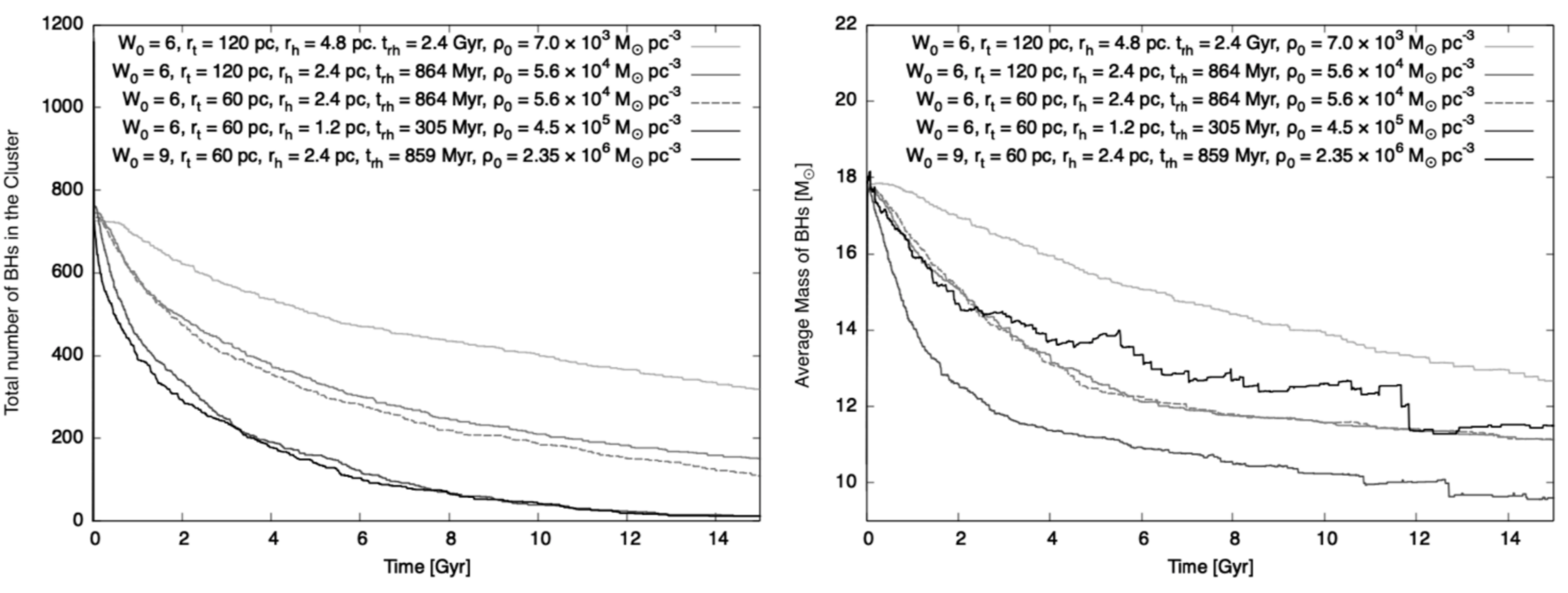} 
 \caption{Evolution of the total number of BHs and average BH mass in GC models with $7\times10^{5}$ objects, 10 per cent initial binaries and metallicity (Z) of  $0.001$. The key indicates the model's initial central concentration, tidal radius, half-mass radius, half-mass relaxation time and central density.}
   \label{fig2}
\end{center}
\end{figure}

\section{Observational Signatures of a Black Hole Subsystem}

\begin{table}
  \begin{center}
  \caption{Galactic GC models that were identified as having a sizeable population of BHs using independent methods in \cite[Askar et al. (2018), Askar et al. (2019) and Arca Sedda et al. (2019)]{2018MNRAS.478.1844A,2019MNRAS.485.5345A,2019arXiv190500902A}. The first column indicates the number of methods by which the GC was shortlisted (\emph{3} - identified in all 3 papers, \emph{2} - identified in 2 of the 3 papers, \emph{1} - identified only in 1 of the 3 papers.)}
  \scriptsize{
\begin{tabular}{cc}
\toprule
\textbf{Identification} & \textbf{Galactic GC Names} \\ \midrule
3 & \begin{tabular}[c]{@{}c@{}}NGC 288, NGC 3201, NGC 4372, IC 4499, NGC 5897, NGC 5986,\\   NGC 6205 (M13), NGC 6712, NGC 6779 (M56), NGC 6809 (M55)\end{tabular} \\ \hline
2 & \begin{tabular}[c]{@{}c@{}}NGC 4833,  NGC 5466, NGC 6101, NGC 6496, NGC 6569, NGC 6584, NGC 6656 (M22), \\ NGC 6723, Pal 1, NGC 5139 ($\omega \ Cen$), NGC 6402 (M14)\end{tabular} \\ \hline
1 & \begin{tabular}[c]{@{}c@{}}NGC 4590, NGC 5272, NGC 6144, NGC 6171, NGC 6362, NGC 6401,\\  NGC 6426, IC 1276,NGC 6934, NGC 6981, NGC 6254 (M10), NGC 1261 NGC 2419,\\  Pal 4, NGC 5053, NGC 6229, NGC 6218 MGC 7006\end{tabular} \\ \bottomrule
\end{tabular}
  }
 \end{center}
    \label{table1}
\end{table}

Investigating the observational properties of the 160 models that had more than 15 BHs at 12 Gyr, we found that the presence of a significant number of BHs prevents segregation of luminous stars in the GC center. Thus, the central surface brightness is lower in models with a BHS as opposed to models that lack black holes. BHS models are also characterized by typically large core and half-light radii. By building on the results of \cite[Arca Sedda et al. (2018)]{2018MNRAS.479.4652A}, \cite[Askar et al. (2018)]{2018MNRAS.478.1844A} we're able to find a correlation between the average surface brightness and the density of BHs inside the BHS. By comparing available observational data \cite[(Harris 1996, updated 2010)]{{1996AJ....112.1487H}} for Galactic GCs with simulated GC models, \cite[Askar et al. 2018]{2018MNRAS.478.1844A} were able to identify 29 Galactic GCs that were most likely to have a substantial number of BHs contained within them. They were also able to estimate the number of BHs in these clusters using the correlations from the simulated GC models.

\cite[Askar et al. (2019)]{2019MNRAS.485.5345A} used supervised machine learning (ML), training on 12 Gyr observational properties from MOCCA-Survey Database I in order to identify if a GC could contain a BHS. These ML techniques were used with parameters for Galactic GCs from the Harris (1996, updated 2010) and \cite[Baumgardt \& Hilker (2018)]{2018MNRAS.478.1520B} catalogues. They identified 18 Galactic GCs that had observed properties consistent with the presence of a large number of BHs. \cite[Arca Sedda et al. 2019]{2019arXiv190500902A} used a multidimensional method to compare 12 Gyr properties of simulated MOCCA GC models with Harris catalogue data to identify which Galactic GCs contain an intermediate-mass BH or a BHS. This approach found 22 Galactic GCs with properties that are best fit by GC models that contain a BHS. In Table 1, we show the Galactic GCs that were identified using these methods. We find that 15-20 per cent of Galactic GCs could contain a significant number of BHs.

\section{Acknowledgements}

AA is supported by the Carl Tryggers Foundation through the grant CTS 17:113. MG and AL are supported by NCN, Poland, through the grant UMO-2016/23/B/ST9/02732.
MAS acknowledges support from the Alexander von Humboldt Foundation, MP received funding from the Marie Skłodowska-Curie grant agreement no. 664931.


\end{document}